\documentclass[12pt,preprint]{aastex}

\shorttitle{Planetesimal Accretion in Binary Systems: the Effects of
Gas dissipation} \shortauthors{Xie & Zhou}
\begin{document}

\title{PLANETESIMAL ACCRETION IN BINARY SYSTEMS: THE EFFECTS OF GAS DISSIPATION  }
\author{Ji-Wei Xie and Ji-Lin Zhou}
\affil{Department of Astronomy, Nanjing University, Nanjing 210093,
China} \email{xjw0809@163.com}

\begin{abstract}
Currently, one of major problems concerning planet formation theory
in close binary systems is, the strong perturbation from the
companion star can increase relative velocities ($\triangle V$) of
planetesimals around the primary and thus hinder their growth.
According to previous studies, while gas drag can reduce the
$\triangle V$ between bodies of the same sizes by forcing orbital
alignment to planetesimals, it increases the $\triangle V$ among
bodies of different sizes. In this paper, focusing on the $\gamma$
Cephei binary system, we propose a mechanism that can overcome this
difficulty. We show that in a dissipating gas disk (with a typical
dissipating timescale of $\sim 10^5-10^6$ years), all the
planetesimals eventually converge towards the same forced orbits
regardless of their sizes, leading to much lower impact velocities
among them. These $\triangle V$ decrease processes progressively
increase net mass accretion and even trigger runaway growth for
large bodies (radius $>15$ km).  The effect of size distribution of
planetesimals is discussed, and found to be one of the dominant
factors that determine the outcome of collisional evolution. Anyway,
it can be concluded that by including the gas dissipation in the
early stage of disk evolution, the conditions for planetesimal
accretion become much better, and the process from planetesimal to
planet-embryo can be carried out in close binary systems like
$\gamma$ Cephei.
\end{abstract}


\section{INTRODUCTION}
With the increasing number of discovered planets in binary systems
and the belief that a majority of solar-type stars reside in binary
or multiple systems, problem of planetary formation in binary
systems becomes a crucial one. Most of discovered planet-bearing
binary systems are S-type systems (e.g. $\gamma$ Cephei system, see
Hatzes et al. 2003) in which planets orbit the primary star with a
companion star surrounding them on an outer orbit. According to the
classical planetary formation scenario, planets form in a
protoplanetary disk of gas and dust orbiting a protostar. The
formation process is usually treated in three stages (Lissauer 1993;
Papaloizou \& Terquem 2006; Armitage 2007): [S1.] formation of
kilometer-size plantesimals ($10^{18}-10^{22}$ g) from sticking
collisions of dust (Weidenschilling \& Cuzzi 1993) or from
gravitational fragmentation of a dense particle sub-disk near the
midplane of the protoplanetary disk (Goldreich \& Ward 1973) on
timescales of the order of $ 10^4$ years, [S2.] accretion of
plantesimals into planetary embryos ($10^{26}-10^{27}$ g, Mercury-
to Mars-size) through a phase of ``runaway" and ``oligarchic" growth
on a timescale of the order of $10^4 - 10^5$ years, depending on
initial planetesimal sizes, duration of the runaway growth period,
possible transition to oligarchic mode (Greenberg et al. 1978;
Wetherill \& Stewart 1989; Barge \& Pellat 1993; Kokubo \& Ida 1996,
1998, 2000; Rafikov 2003, 2004). [S3.] giant impacts between
embryos, producing full-size ($10^{27}$ to $10^{28}$ g) terrestrial
planets in about $10^7-10^8$ years (Chambers \& Wetherill 1998;
Kokubo, Kominami \& Ida 2006; Levison \& Agnor 2003). Here we focus
on the stage II to see the influence of the companion on the
planetesimal accretion.

The companion star, especially when it is on a close orbit with a
high eccentricity, may prevent planetary formation through reducing
the size of the accretion disk (Artymowicz \& Lubow 1994), and
exciting high relative velocities between colliding planetesimals
(Heppenheimer 1978; Whitmire et al. 1998). The relative velocity
($\triangle V$) is a critical parameter, which determines whether
accretion or erosion dominates. Due to the perturbation by the
companion, $\triangle V$ may exceed the planetesimal escape velocity
($V_{esc}\sim100\times{(R_{p}/100km)}$ m s$^{-1}$), and thus inhibit
runaway growth. Furthermore, $\triangle V$ can even exceed the
threshold velocity ($V_{ero}$) for which erosion dominates
accretion. Here $V_{ero}$ is a few times larger than $V_{esc}$,
depending on the prescription on collision.

Since planetesimals orbit the star in a sub-Keplerian gas disk
(Adachi et al. 1976), the presence of gas drag does not only damp
the companion's secular perturbation, but it also forces a strong
periastron alignment of planetesimal orbits. This alignment
significantly reduces $\triangle V$ between equal-sized bodies, favoring the accretion process
(Marzari \& Scholl 2000). Nevertheless, the alignment forced by the
gas drag induces another problem.  As the alignment is
size-dependent, it can only reduce $\triangle V$ between
planetesimals of the same sizes, and at the same time it increases
$\triangle V$ between planetesimals of different sizes. Thebault et
al. (2006) find that this differential orbital alignment is very
efficient, leading to a significant $\triangle V$ increase for any
departure from the exact equal-size condition ($R_{1}$ = $R_{2}$,
where $R_{1}$ and $R_{2}$ are the radiuses of the two colliding
bodies).

Pervious studies adopted a steady gas disk in which dissipating
process was neglected and the local gas density was constant. This
assumption, which is valid only when the planetesimal accretion time
scale (of the order of $10^{4}$ to $10^{5}$ years) is much shorter than the dissipating
time scale of local gas density, is violated under the following
conditions. 1)When the disk viscosity is high or photoevaporation
from external star exists (Hollenbach et al. 1994, 2000; Matsuyama
et al. 2003), disks can dissipate very fast and have short lifetimes
within a few $10^{5}$ years. 2)It is suggested that the assumption
of a single time scale for disk dissipation is not correct, and there could be a wide spread of disk lifetimes, with a large fraction of short-lived disks (Bouwman et al. 2006). As calculated by
Matsuyama et al.(2003) and Alexander et al. (2006b), even for a disk
with a lifetime of the order of $10^{6}$ years, local density can
decrease by as many as two orders of magnitude within the first few
$10^{5}$ years. For these two considerations, therefore, a model
that includes gas dissipation is essential for studying planetesimal
accretion.

In this paper, we consider a model in which gas density
progressively decreases, to see how the conditions of planetesimal
accretion are affected by the gas dissipating process. As expected,
the planetesimal growth conditions change to being
accretion-friendly due to an dissipation induced orbital
convergence, which reduces $\triangle V$ between bodies of different
sizes. We describe our numerical model and methods in section 2. In
section 3, first, we simply review the planetesimal dynamics under
the coupled influence of secular perturbation and gas drag, and then
present the results. Some related and crucial issues, such as the
radial drift, impact rate, erosion conditions and remanent gas, are
discussed in section 4. Finally, in section 5, we summarize this
paper.

\section{NUMERICAL MODEL AND METHODS}
\subsection{Gas Disk Model}
We made the gas model similar to that of Thebault et al. (2004).
Following Weidenschilling and Davis (1985), the gas drag
 can be expressed as:
\begin{equation}
   \mathbf{F}={-Kv\mathbf{v}} ,
\end{equation}
where $\mathbf{F}$ is the force per unit mass, $\mathbf{v}$ the
relative velocity between the planetesimal and gas, $v$ the velocity
modulus, and $K$ is the drag parameter defined as:
\begin{equation}
   K={ {3\rho_{g}C_{d}} \over {8\rho_{p} R_{p}} },
\end{equation}

\begin{equation}
\rho_{g}=\rho_{g0}T^{-n}, T={t\over T_{s}+1},
\end{equation}
where $\rho_{g}$ is the local gas density with an initial value of
$\rho_{g0}$ , $\rho_{p}$ and $R_{p}$ the planetesimal density and
radius, respectively. $C_{d}$ is a dimensionless coefficient related
to the shape of the body ($\simeq 0.4$ for spherical bodies). The
$T^{-n}$ function, in which time $T$ is scaled by $T_{s}$, is used
to include the gas dissipation, and it is based on the analytic
similar solutions given by Lynden-Bell and Pringle (1974). Taking
typical parameters from Hartmann et al. (1998), where $n=3/2$,
$T_{s}=10^{5}$ years, we plot figure 1 to show the gas disk density
evolution vs. time. The gas disk is scaled by the Minimum Mass solar
Nebula (hereafter MMN for short) and has the same profile to the
MMN(Hayashi 1981). The initial gas density is 10 MMN, and the
corresponding disk mass is about 100 Jupiter mass. As shown in
figure 1, gas density rapidly decreases from 10 MMN to 0.5 MMN
within the first few $10^{5}$ years, and then it experiences a slow
damping process lasting for a few million years. This dissipation
model is consistent with current theoretic calculations (Matsuyama
et al. 2003; Alexander et al. 2006a, 2006b) and observations (Strom
et al. 1993; Haisch et al. 2001; Chen \& Kamp 2004), which suggest a
typical disk age of 1 million years with a large scatter from $0.1$
to $10$ million years. Notice that the effects of binarity on the
dissipation of gas disk are not taken into account because details
of these issues are poorly known at present.

Our model implicitly assumes an axisymmetric gas disk with constant
circular streamlines and follows a classical Hayashi (1981) power
law distribution. We are aware that this is a crude simplification
for modeling gas disk in close binary systems. In reality, the gas
disk around the primary also ``feels" the companion's perturbation,
under which disk structure would vary from the simplified gas model.
For example, the companion's perturbation can induce spiral
structures within the disk(Artymowicz and Lubow, 1994). To fully
model the behavior of planetesimals in these complex gas disks, one
would probably have to rely on hydro-code modeling of the gas in
addition to N-body type models for planetesimals. Such an
all-encompassing gas plus planetesimals modeling goes beyond the
scope of our study in this paper, and it is certainly the direction
of further binary disk studies. Therefore, taking a first step here,
we just prefer a simplified approach where gas drag force is given
by equation (1). As discussed by some previous studies (Scholl et
al. 2007, Thebault et al. 2006), this kind of simplification, on the
average, is reasonable at least for the dynamical evolution of
kilometer-size planetesimals.

\subsection{Initial Conditions}
We focus on the $\gamma$ Cephei system, which is a close S-type
binary planetary system, hence being a good example to test the
influence of the companion on planetesimal accretion. Most
parameters adopted in this paper are listed in table 1. The initial
gas disk has the same profile to MMN, but is denser by 10 times. We
concentrate on planetesimals of four radiuses ($R_{p}=2.5, 5, 15,
50$ km). As stated by Thebault et al. (2006), for impacts between
small bodies(1 $< R_{p} <$ 10 km), the delivered kinetic energy
peaks at roughly $R_{1}\simeq1/2R_{2}$, where $R_{1}$ and $R_{2}$
are the radiuses of the two colliding bodies. For the bigger ones,
this $R_{1}/ R_{2}$ ratio is somewhat smaller. Hence, the relative
velocity $\triangle V(2.5,5)$ between bodies of $R_{p}=2.5$ km and
$R_{p}=5$ km can be typical example values for small planetesimals,
and $\triangle V(15,50)$ for large ones.  All the planetesimals
initially have very small inclinations based on the work of Hale
(1994), which suggests that approximate coplanarity between the
equatorial and orbital planes exist for solar-type binary systems
with separations less than 30-40 AU. Since it is unrealistic that
all planetesimals form synchronously, some earlier formed
planetesimals may have been pumped up to eccentric orbits while some
others have just formed. For this reason, the initial planetesimal
orbits should have random eccentricities within the range from 0 to
$e_{max}$, where $e_{max}$ is the maximum eccentricity that pumped
up by the companion. In the $\gamma$ Cephei system, $e_{max}$ is
about $0.1$ at $2$ AU from the primary.

One implicit initial condition in this paper is that, of course,
kilometer-size plametesimals have already formed when the disk
begins dissipating. At present, with the poor knowledge on
planetesimal formation in binary systems, whether this assumption is
valid or not is not for sure at all. According to current limited
knowledge on planetesimal formation around a single star,
kilometer-size planetesimal can form within $10^{3} - 10^{5}$ years
through sticking collision or by gravitational instability after
dust having settled down on the midplane (Lissauer 1993;
Weidenschilling 1997; Goldreich \& Ward 1973; Youdin \& Shu 2002).
In such case , the timescale of planetesimal formation can be much
shorter than that of gas disk dissipation (about $10^6$ years is
considered in this paper), and thus it is reasonable to assume that
the gas dissipation starts when a population of kilometer-size
planetesimals exist in the system.

\subsection{Numerical Methords}
We performed two kinds of runs. First, we numerically integrated the
equations of motion for $1000$ independent planetesimals with
semi-major axes from $1$ to $4$ AU. The focus is put on the
time-evolution of orbital eccentricities and of orbital periastrons.
Second, we concentrate on the time-evolution of $\triangle V$ at a
specific region near $2$ AU from the primary star where a planet is
detected. This is the configuration of the $\gamma$ Cephei system
that we specifically consider here. Plantesimals are initially
distributed in a ring near $2$ AU. Since the planetesimal sizes
(order of km) are very small comparing to the system typical scale
(order of AU), it is very difficult to track all ``real" physical
impacts among these planetesimals (Brahic 1977, Charnoz et al.
2001a, 2001b; Lithwick \& Chiang 2007, etc). In such case, we have
to resort to the classical ``inflated radius" assumption, which
assumes an artificially increased radius to each particle (e.g.
Brahic 1977; Thebault \& Brahic 1998; Marzari \& Scholl 2000). For
planetesimals considered here, an artificially increased radius
(about $10^{-5}-10^{-4}$ AU) of $100$ times larger than the ``real"
radius is adopted for each planetesimal.

In all the runs, we used the fourth order Hermite integrator (Kokubo
et al. 1998), including the gas drag force and the perturbation of
companion. As gas drag also forces inward drift of planetesimals, we
adopt following boundary conditions: bodies whose  semi-major axes
are less than $R_{in}$(greater than $R_{out}$), will be reset to
$R_{out}$($R_{in}$), where $R_{in}$ and $R_{out}$ are the inner and
outer boundaries of planetesimal belt, respectively. In these
resetting processes, only the semi-major axes of those bodies are
changed, while other orbital elements are preserved.

\section{RESULTS}
\subsection{Planetesimal Dynamics: the Secular Approximation}
Before presenting the results, let's review the planetesimal
dynamics in a perturbed system. Heppenheimer (1978) developed a
simplified theory for the evolution of planetesimal eccentricity
with time in binary systems. First, he defined two variables $h$ and
$k$ as
\begin{equation}
   h=e_{p}  sin(\varpi),  k=e_{p}  cos(\varpi) ,
\end{equation}
where $e_{p}$ is the planetesimal eccentricity and $\varpi $ its
periastron longitude defined with respect to that of the companion
star ($\varpi=\varpi_{p}-\varpi_{B}$, where $\varpi_{p}$ and
$\varpi_{B}$ are the periastron longitudes of the companion and the
planetesimal, respectively. Then, introducing in the Langrange
planetary equations, he obtained the following equations for $h$ and
$k$:
\begin{equation}
   {{dh}\over{dt}} =Ak - B ,
\end{equation}

\begin{equation}
   {{dk}\over{dt}} =- Ah ,
\end{equation}
where the constants $A$, $B$ are
\begin{equation}
   A = {3\over4} {M_{A} \over {n(1-e_{B}^{3/2})}},
   B = {15\over16} {ae_{B}\over{n(1-e_{B}^{5/2})}},
\end{equation}
with $e_{B}$ the eccentricity of the binary system and $M_{A}$ the
mass of the primary star. $a$ and $n$ are the semi-major axis and
mean motion of the planetesimal, respectively. The units of mass,
distance, and time are normalized in such a way that the
gravitational constant $G$ and the sum of the masses of the two
stars are set equal to $1$. The semimajor axis of the binary $a_{B}$
is chosen as the units of length, so that the time is expressed in
units of $(1/2\pi)T_{B}$, where $T_{B}$ is the orbital period of the
binary system.

In $h-k$ plane, there is an equilibrium point (where $dh/dt=0,
dk/dt=0$) for equations (5) and (6), which is referred to as E0 in
this paper. At E0, $e_{p}=e_{f}$ and
$\varpi_{p}=\varpi_{f}=\varpi_{B}$, where $e_{f}=B/A$ and
$\varpi_{f}=\varpi_{B}$ are the forced eccentricity and periastron
of the planetesimal respectively. If a planetesimal reaches the
equilibrium point E0, its eccentricity and periastron will fix on
$B/A$ and $\varpi_{B}$ forever.

To compute the effect of gas drag on the variables $h$ and $k$,
Marzari and Scholl (2000) modified equations (5) and (6) as
following:
\begin{equation}
   {{dh}\over{dt}} =Ak - B - Dh(h^2 + k^2)^{1/2},
\end{equation}

\begin{equation}
   {{dk}\over{dt}} =- Ah - Dk(h^2 + k^2)^{1/2},
\end{equation}
where $D$ is a coefficient to measure the gas drag force. According
to these equations, for a specific $D$, the planetesimal orbit will
quickly or slowly (depending on the $D$ value, a larger value $D$
leads to a faster speed) reaches another equilibrium point(different
with E0), with an equilibrium eccentricity below $B/A$. Furthermore,  AND THIS IS THE CRUCIAL POINT OF THIS STUDY, if $D$ damps slowly(caused by gas dissipation),
planetesimals will shift their orbits from this equilibrium point
eventually toward E0.

In figure 2, we illustrate these processes. In no gas case, the
motion in the $h-k$ plane is circulating around the equilibrium
point $E0$ that derived from equations (5) and (6). For the case
with gas drag in our gas disk model, motions are divided into the
following two phases: a) ``no dissipation phase" in the first few
$10^3$ years, in which gas disk dose not significantly dissipate and
planetesimals of different sizes quickly reach different equilibrium
points depending on their sizes(point $E1$ for bodies of 50 km, $E2$
for 20 km, see figure 2), b) ``dissipating phase", in which gas disk
gradually dissipates, at the same time all the motions shift along
the line $E4-E0$, and eventually fix on the same equilibrium point
$E0$ regardless of their sizes. We also analyze the effects of
initial orbits on the dynamical behavior. As shown in figure 2,
bodies with the same sizes(5 km) but different initial orbits (one
is at $I1$, the other is at $I2$) go through different paths($I1-E4,
I2-E4$) to reach the same equilibrium point($E4$). After that, they
both experience the same ``dissipating phase" from $E4$ to $E0$.
From this point, we can see that how to choose the initial
planetesimal orbits do not affect the final results which are based
primarily on the latter ``dissipating phase".

The appearance of the dissipating phase and the dynamical behavior
of planetesimal orbits during this phase are very important because
they provide channels to reduce the differential phasing effect
induced by the size-dependence of gas drag. Based on the above
theoretical analysis, we can expect a relative velocity($\triangle
V$) decrease from the convergence of all the planetesimal orbits. In
the next two subsections, we will numerically simulate this
$\triangle V$ decrease process.

\subsection{Time-evolution of Eccentricity and Periastron}
  We first performed a simulation in which 1000 planetesimals (4
equal-number groups: $R_{p}=2.5, 5, 15, 50$ km, mutual interactions
were neglected) were initially distributed between 1 AU and 4 AU
from the primary. Figure 3 shows the distributions of planetesimal
eccentricities and periastrons vs. semi-major axes at different
epoches. Beyond 3 AU, the distributions of planetesimal
eccentricities and periastrons are random because the shorter period
perturbation and mean motion resonances are dominant there. Thus,
hereafter only planetesimals within 3 AU are discussed. In figure
3a(or b), every eccentricity (or periastron) reaches an equilibrium
value at 5,000 years. These equilibrium values, as discussed in the
above subsection and also pointed out by previous studies(Thebault
et al. 2006), depend on the balance between the perturbation by the
companion and the gas drag force. Due to the size-dependence of gas
drag force, bodies of different sizes reach different equilibrium
eccentricities (or periastrons). The four lines in each panel are
corresponding to bodies of four kinds of sizes ($R_{p}=2.5, 5, 15,
50$ km). As the gas dissipates gradually, the equilibrium
eccentricities (or periastrons) move to larger values, but at the
same time the differences among them become smaller (see Fig.~3c(or
d)). After a long time (5,000,000 years, see Fig.~3e(or f)), almost
all eccentricities (or periastrons) converge towards $e_{f}$
($\varpi_{f}$).

\subsection{Time-evolution of Relative Velocity}
We perform another simulation to investigate the time-evolution of
$\triangle V$ in a specific place(at 2 AU from the primary). In this
calculation, 1000 Planetesimals were initially distributed with
major-axes between 1.5 and 3 AU. This planetesimal ring is wide
enough that to trace most of collisions at 2AU.

The results are plotted in figure 4.  Figure 4b and figure 4c show
the average eccentricity and periastron of bodies at 2 AU as the
functions of time, respectively. As disk gradually dissipates, all
the planetesimals converge towards the same forced orbits where
$e_{p}=e_{f},\varpi_{p}=\varpi_{B}$(also see E0 in figure 2). Figure
4a plots the $\triangle V(R_{1},R_{2}$) as the function of time. It
is evident that the larger differences in orbital elements, the
larger value of $\triangle V$. From figure 4a, it appears that the
$\triangle V$ between bodies of equal-size are always small because
of the orbital alignment. However, the $\triangle V$ between bodies
of different sizes first increase quickly to high values (e.g.
$300\sim800$ m s$^{-1}$ ), then each of them experiences a
relatively slow decrease.  This $\triangle V$ decrease is most
efficient for large bodies. For 15 km-size and 50 km-size bodies,
the relative velocity $\triangle V(15,50)\sim300$ m s$^{-1}$  is
much larger than their escape velocities $V_{esc}\sim$ 50 m s$^{-1}$
at the beginning. After about $3\times10^{5}$ years, $\triangle
V(15,50)$ get lower(about 40 m s$^{-1}$) than the escape velocities
of the large planetesimals, so that runaway growth can occur.

To compare with the dissipating gas drag case showed in figure 4, we
perform one more case with constant gas drag.  It shows, in figure
5, that without gas dissipation every $\triangle V$ is forced on a
relatively high value determined by the equilibrium between the gas
drag force and secular perturbation. The main difference with the
dissipating gas case is that there is no late stage with
size-independent orbital phasing and thus no $\triangle V$ decrease.

\section{DISCUSSIONS}

\subsection{Impact Rate}
As impacts of different types (between the same sizes or different
sizes) have totally different $\triangle V$ and thus different
outcomes (erosion, incomplete accretion, complete accretion and
runaway growth), the condition that which type of collision
dominates becomes crucial for planetesimal growth. Figure 6 plots
the distributions of impact rates for two cases: a)standard case,
b)random case. In both cases, we compute 1000 planetesomals whose
radius distribution is assumed as a gaussian function centered at 8
km with a dispersion $\triangle R=7$ km. The only difference between
them is the companion and gas drag are not included in the random
case. As shown in figure 6, for the random case, the distribution of
impact rates depends only on the initial size distribution: impacts
occur more often in the places where more planetesimals are
distributed for impacts between equal-sized bodies close to the center of the Gaussian. On the other hand, in the standard case, the
distribution is obviously size-dependent: impacts mainly occur
between bodies of different sizes. By comparing these two cases, it
is clear: under the coupled effect between gas drag and the
companion's perturbation, impacts between bodies of different sizes
are favored, while impacts between bodies of the same(or similar)
sizes are hindered. This result can be understood in this way: for
bodies of the same sizes, as they have the same forced orbits and
radial drifts, one can only collide with another when their
semimajors are very close; for bodies of different sizes, in
contrast, as they have different forced orbits and radial drifts,
one can cross many more planetesimal orbits on a much larger region.

\subsection{Accretion or Erosion}
The key result of this paper presented in section 3 is: as gas
dissipates, all planetesimals eventually converge towards the same
forced orbits regardless of their sizes, leading to much lower
$\triangle V$ than in the constant-gas density case. To further see
the effects of these $\triangle V$ decreasing processes on
planetesimal collisional evolution(accretion or erosion), we then
perform a quantitative study.

Following Kortenkamp \& Wetherill (2000), we adopt the disruption
limit given by Love and Ahrens (1996), and compute the net mass
accretion ratio (see Appendix for details) for every impact. Figure
7 shows the time-evolution of net mass accretion ratios($A_{r}$) for
impacts between different size groups. For impacts between bodies of
the same sizes, net mass accretion ratios are not plotted, since the
$\triangle V$ are always low enough for runaway growth in such
cases. As shown in figure 7, it can be summarized as following: 1)
for small bodies ($R_{p}<5$ km), collisions always lead to erosion
during the first $7\times 10^5$ years, after which accretion occurs
with a progressively increasing $A_{r}$, 2) for intermediate
bodies($5<R_{p}<15$ km), $A_{r}$ is initially modest(75\%-80\%) and
will progressively increase (to 90\%-95\%) as the gas dissipates, 3)
for large bodies ($R_{p}>15$ km ), $A_{r}$ is always very
high($\geq95\%$), 4) for impact between a large($R_{p}>15$) km and a
small($R_{p}<5$ km) planetesimal, while the $\triangle V$ is high
and decreases slowly(see figure 4), $A_{r}$ is always high ($\geq
95\%$). Therefore, to fully know the details of collisions among a
swarm of planetesimals will have to require an entire information of
the initial planetesimal size distribution, which is, however, not
clear at all with current knowledge.

Here, for simplicity, we just perform four simplified tests assuming
for the planetesimal size distribution a gaussian and three power-law
functions, respectively. For the three power law cases, planetesimals have
distributions given by $N\propto m^{-1.7}$ (Makino et al. 1998) with
three radius ranges, namely 1 - 50 km, 2.5 - 50 km and 5 - 50 km. For the gaussian case, the radius distribution is assumed as the gaussian function centered at
8 km with a dispersion $\triangle R=7$ km. Figure 8 plots the
time-evolution of the average $\triangle V$ and $A_{r}$ for these
four cases. 
It shows, at the first few $10^3$ years(no dissipation
phase), the conditions for accretion or erosion totally depend on
the initial size distribution of planetesimals. In this phase, the
average $\triangle V$ is pumped up by the size-dependence of orbital
alignment, and thus the accretion is inefficient($A_{r}\sim75\%$)
for one power law case(5 - 50 km), dangerous ($A_{r}\sim30\%$) for
the gaussian case and  another power law case(2.5 - 50 km), and even completely suppressed for the power law case(1 - 50 km)  . However, after a few $10^5$ years(gas dissipation phase), all the $\triangle V$ get low enough and accretion is
efficient($A_{r}\geq 95\%$) for all the cases, regardless of the
initial size distribution of planetesimals. Notice that the smaller bodies we consider initially, the more time the system needs to become accretion-friendly. For the power law case with minimum size of 1 km, it indeed takes about $6\times10^5$ years before accretion is efficient. As discussed in the next subsection, this long timespan can worsen the radial drift problem.

\subsection{Radial Drift}
Moving in the gas disk, planetesimals undergo a headwind by which
they are forced to progressively migrate inwards (Adachi et al.
1976). In the above runs, we adopt a boundary condition described in
section 2.3 to keep all the planetesimals staying in our computing
zone ($1.5 - 3$ AU). This is reasonable only if the planetesimal
disk is extended enough so that planetesimals can flow into the
computing zone from the outer disk. However, theoretical
calculations of binary-disk interactions predict that companions
might truncate circumstellar disks at an out radius of $0.2 - 0.5$
times the binary semi-major axes (Artymowicz \& Lubow 1994). For
$\gamma$ Cephei system, $a_{B}=18.5$ AU, then the truncated disk
size is about $3.7-9.3$ AU. Therefore, there may be not enough
material supplied from the outer disk,  and it means there should be
enough planetesimals staying in the computing zone for at least a
few $10^{5}$ years to form planets. For this reason, we performed a
simulation without any boundary condition to compare the results in
figure 4. We find most large bodies with $R_{p}=15$ km and
$R_{p}=50$ km stay in the computing zone, having $\triangle V$
curves similar to those in figure 4, while almost all the small
bodies with sizes of $R_{p}=2.5$ km and $R_{p}=5$ km are removed by
gas drag. This problem of ``too fast migration" will be even worse
when smaller bodies are considered, such as bodies with radiuses of
$1-10$ m.  As shown in figure 8, for the power law case(1 - 50 km), there is $6\times10^5$ years timespan, during which erosion dominates and thus planetesimals are transformed into small fragments which are quickly removed by inward drift.

Actually, fast inward drift induced by gas drag is a general problem in the classical
planet formation model(Lissauer 1993; Papaloizou \& Terquem 2006;
Armitage 2007), and several ways have been proposed to address this
issue. It is possible that large planetesimals ($R_{p}>10$ km, which is big enough to overcome the inward dirft) form directly via
gravitational instability in a few $10^3$ years (Goldreich \& Ward
1973; Youdin \& Shu 2002). In addition, radial drift may allow small
bodies to pileup within the inner disk to form larger planetesimals(
Youdin \& Chiang 2004), and the present of turbulence in gas disk
can also reduce the radial drift(Durisen et al. 2005; Haghighipour
\& Boss 2003; Rice et al. 2004).

\subsection{Remanent Gas for gaseous Planet Formation}
In addition, there should be enough remanent gas to form a massive
gaseous planet, as required to fit the minimum mass ($\sim2$ jupiter
masses) of the planet detected in the $\gamma$ Cephei system. In
 this paper, for a initial gas disk of 10 MMN(about 100 Jupiter
mass), after $5 \times 10^{5}$ years when most $\triangle V$ have
already decreased to low enough values, the remanent gas, according
to figure 1, is about 7 Jupiter masses.  On the other hand, Kley and
Nelson (2007) suggest that the gas accretion onto a planet will be
highly efficient in the $\gamma$ Cephei system due to the large
induced planet orbital eccentricity. Their simulations indicate that
it needs a gas disk with only $\sim 3$ Jupiter masses to form a
gaseous planet of $\sim 2$ Jupiter masses. Therefore, it is possible
to form a massive gaseous planet in our dissipating gas model.

\section{SUMMARY}
In this paper, focusing on the $\gamma$ Cephei system and
concentrating on planetesimal impact velocities($\triangle V$), we
numerically investigate the conditions for planetesimal accretion in
binary systems. We extend the studies of Thebault et al. (2004,
2006) by including the effect of a dissipating gas disk.  We confirm
some of their results that in a gas disk without dissipation,
differential orbital alignment is very efficient and increase
$\triangle V$ between bodies of different sizes to high values that
significantly inhabit planetesimal growth. Furthermore, we find that
by including gas dissipation, the differential phasing effect
induced by the size-dependence of gas drag can be reduced. In such
case, as gas density decreases, all planetesimals converge their
orbits towards the same forced orbits, regardless of their sizes.
This orbital convergence induced by gas dissipation is most
efficient for large bodies(15 - 50 km). Within $3\times10^5$ years,
$\triangle V(15,50)$ decrease to low enough values(about 40 m
s$^{-1}$ below the escape velocities of large bodies) for which
runaway growth is able to occur.

In order to get more information of the collisional evolution, we
first discuss the impact rate distribution. We find, for binary
systems including gas drag, collisions between bodies of different
sizes are dominant due to the differential orbital alignment and the
size-dependence of the radial drift.
Considering this result, our mechanism which can reduce the
$\triangle V$ between bodies of different sizes, therefore, becomes
much more essential for planetesimal growth.

By defining the net mass accretion ratio($A_{r}$), we then discuss
the conditions of accretion or erosion for a swam of planetesimals
with different size distributions. We find the size distribution is
a very crucial factor that influences the collisional evolution. For
the constant gas density case, it totally dominates the growth of
planetesimals, and accretions are  only efficient between equal-sized bodies in such
case. On the other hand, for the dissipating gas density case,
effect of size distribution is dominant only at the beginning, and
after a few $10^5$ years, accretion( or even runaway growth) is
always favored, regardless of the initial size distribution of
planetesimals.

Due to the companion's perturbation in a binary system, disk is
truncated to a smaller one and the planetesimals undergo a much
faster inward drift. These effects may induce a problem that whether
enough planetesiamls can remain in the planet-formation zone against
the inward migration. We perform some computations for this
consideration, and find most small bodies($R_{p}<$ 10 km) are
removed within a few $10^5$ years, while no significant influences
on large bodies($R_{p}>$ 15 km). Furthermore, the inward drift problem will be much more acute when the initial planetesimal population is composed mainly of small bodies($R_{p}<$ 2.5 km). In such case, erosion dominates for the first few $10^5$ years, and planetesimals are transformed into small fragments  which are quickly removed by inward drift. 

Finally, we estimate the remanent gas for forming a gaseous planet.
In our dissipating gas disk model, after $5 \times 10^{5}$ when
$\triangle V$ among most of  planetesimals have already decreased to low
enough values, the disk mass is about 7 Jupiter mass which is enough
to form a massive gaseous planet.

\acknowledgments
 We thank  the anonymous referee for valuable suggestions, and
 W. Kley for useful discussions. This work is supported by
 NSFC(10778603), National Basic Research Program of China(2007CB4800).

\appendix
\section{APPENDIX}
\subsection{Net Mass Accretion Ratio} For the sake of simplicity,
colliding planetesimals, both the target and the projectile are
normally considered as nearly homogeneous and spherical bodies, and
all the collisions are treated as central impacts. Having these
assumptions, to describe a specific collision needs only three input
parameters: mass of target($M_{t}$), mass of projectile($M_{P}$),
and impact velocity ($V_{imp}$, namely the $\triangle V$ derived
from our simulations).

Given a target and projectile of mass and radius $M_{t}$,$R_{t}$ and
$M_{p}$, $R_{p}$ respectively, the surface escape velocity of the
pair is
\begin{equation}
   V_{esc}^2={2G(M_{t}+M_{p})\over(R_{t}+R_{p})},
\end{equation}
Where $G$ is the constant of gravity. The center of mass impact
energy available for fragmentation is given by
\begin{equation}
   Q_{f}={k_{1}\over2}V_{imp}^2M_{t}M_{p}/(M_{t}+M_{p}),
\end{equation}
where the impact efficiency $k_{1}=0.5$ is the fraction of the
impact energy not lost to heating. Assuming the crushing strength
scaled by Love and Ahren (1996)
\begin{equation}
   Q_{c}=24.2[R_{t}(cm)]^{1.13},
\end{equation}
where $R_{t}$ is the radius of the target in cm, then the mass of
material fragmented by the impact is
\begin{equation}
   M_{f}=Q_{f}/Q_{c}.
\end{equation}

As some fragments fall back on the target by the gravity, the mass
of material to escape is only a fraction of $M_{f}$ and given by
\begin{equation}
   M_{e}=k_{2}M_{f}V_{esc}^{-2.25}
\end{equation}
(Greenberg et al. 1978), where $k_{2}=3\times10^6$ (cm
 s$^{-1})^{2.25}$.
 Here in this paper, we define a ratio as being
 \begin{equation}
    A_{r}=1-M_{e}/M_{p},
\end{equation}
to measure the fraction of mass accreted on
 the target. If the derived $M_{e}\ge M_{p}$ there is no growth of the target, and $A_{r}=0$ is forced in such cases.
Figure 9 maps the $A_{r}$ in the $R_{1}$-$R_{2}$ plane with four
typical impact velocities: 100m/s, 300m/s, 600m/s, 1000m/s. As shown
in figure 9, bodies with radius below 5 km hardly accrete each
other, on the other hand, once one of the two colliding bodies has
radius larger than 15 km, accretion is always efficient.

\clearpage
\clearpage
\begin{figure}
\begin{center}
\includegraphics[width=\textwidth]{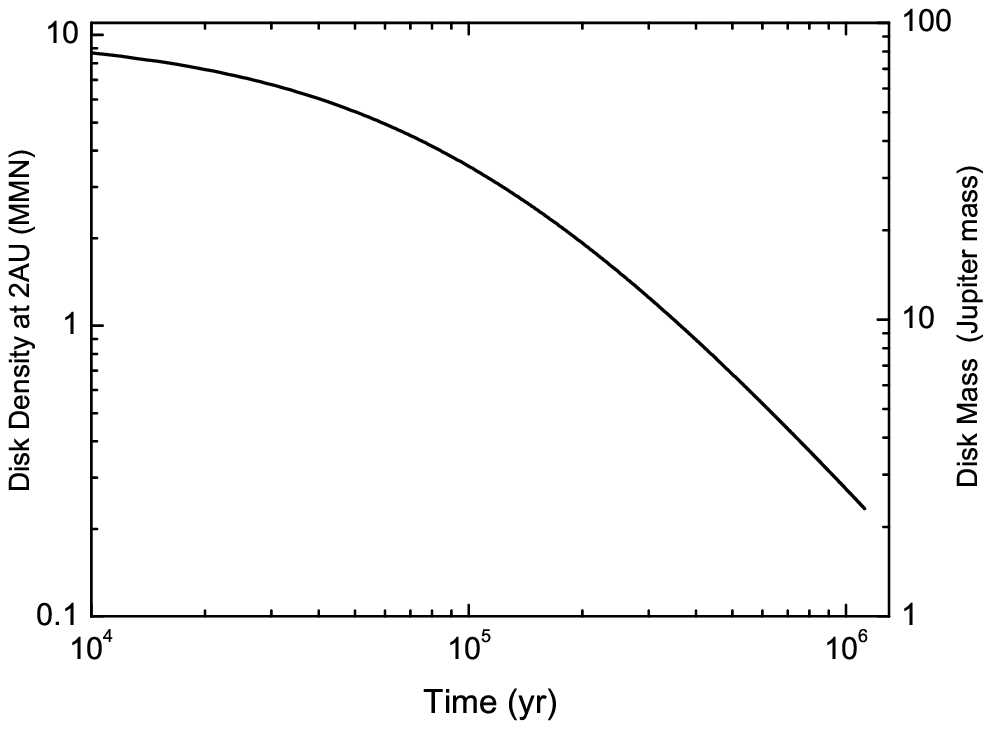}
  \caption{Gas disk dissipation with time. This curve is derived from Eq.(3).
  The left y-axis is the gas density at 2 AU with a initial value $\rho_{g20}=2\times10^{-9}$g cm$^{-3}$(about 10 MMN), and
   the right one is the corresponding disk mass in Jupiter mass units. The mass of 1 MMN disk is estimated as 10 Jupiter mass roughly.}
   \end{center}
\end{figure}

\clearpage
\begin{figure}
\begin{center}
\includegraphics[width=\textwidth]{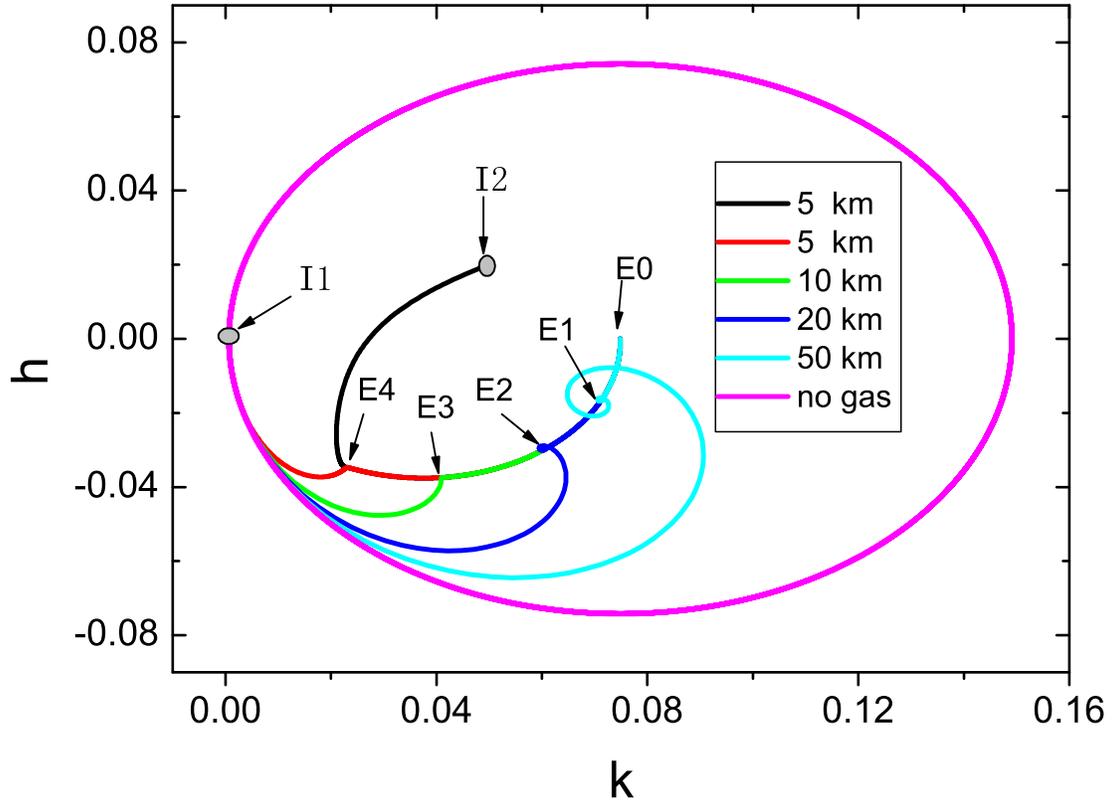}
  \caption{Phase diagram in h-k plane. Gas disks are included for all the cases,
  except for the one denoted by the circle. E0 is the equilibrium point without gas,
  while E1, E2, E3, E4 are the equilibrium points for bodies of 50, 20, 10, 5 km respectively.
  All the motions are initially at point I1 (h=k=0), except for the black one that comes from point I2.}
   \end{center}
\end{figure}

\clearpage
\begin{figure}
\begin{center}
\includegraphics[width=\textwidth]{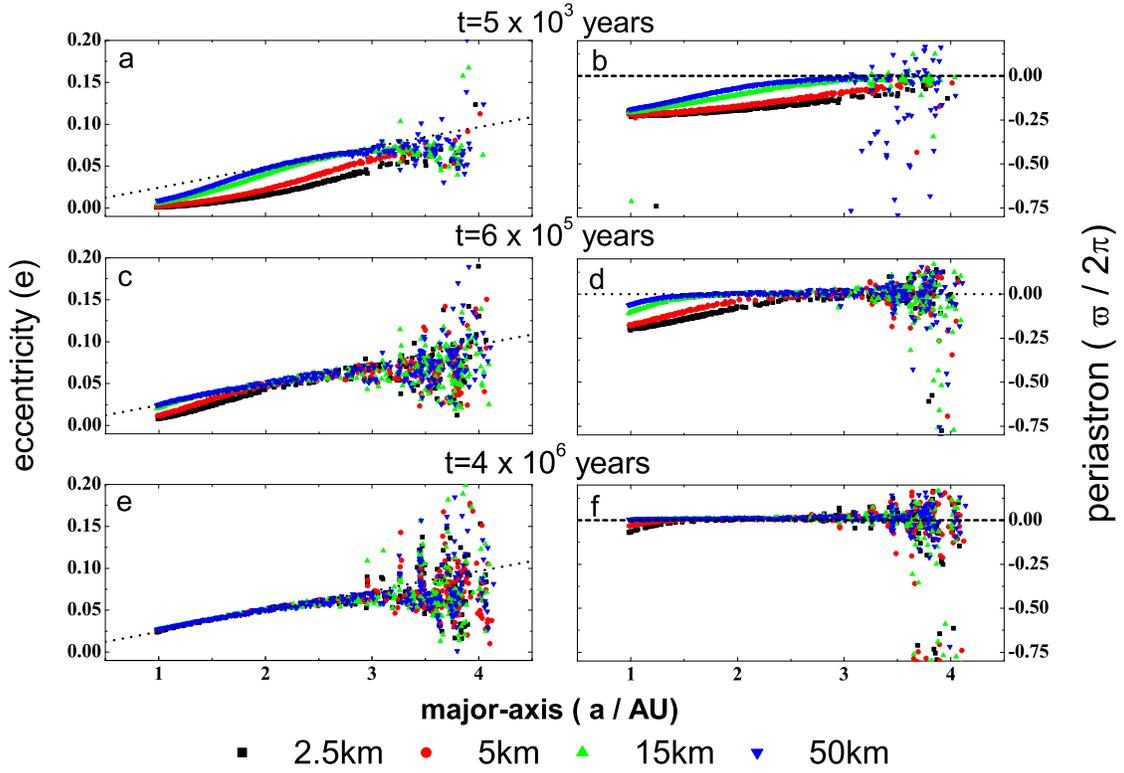}
 \caption{Distributions of planetesimal eccentricities and periastrons vs.
 semi-major axes at three different epoches. Bodies of different sizes
 are ploted in different styles. The dashed lines in the 3 left panels
  and 3 right ones denote $e_{f}$ (forced eccentricity) and
  $\varpi=0$ (which means the planetesimal periastrons $\varpi_{p}=\varpi_{f}$ the forced periastron), respectively.
    }
\end{center}
\end{figure}

\clearpage
\begin{figure}
\begin{center}
\includegraphics[width=\textwidth]{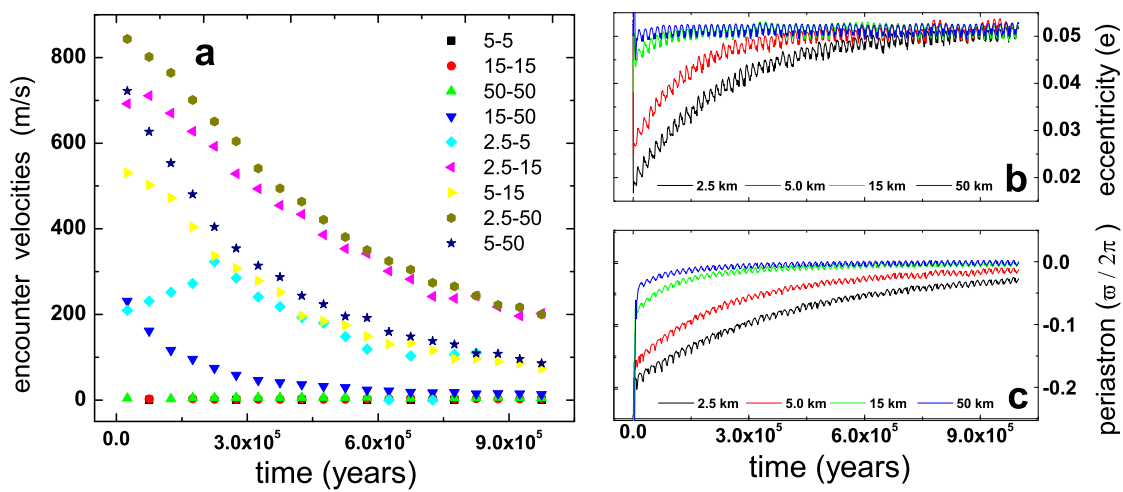}
 \caption{Time-evolutions of $\triangle V$ and distributions of orbital elements at 2 AU from the star.
 (a)Average encounter velocities $\triangle V$ at 2 AU from the primary star
  v.s. time in dissipating gas case. $\triangle V$ between bodies of different sizes are
  plotted in different styles. (b) and (c): Average eccentricity and periastron($\varpi$) at
  2 AU from the primary v.s. time in dissipating gas case. Bodies of different sizes
 are plotted in different styles  }
\end{center}
\end{figure}

\clearpage
\begin{figure}
\includegraphics[width=\textwidth]{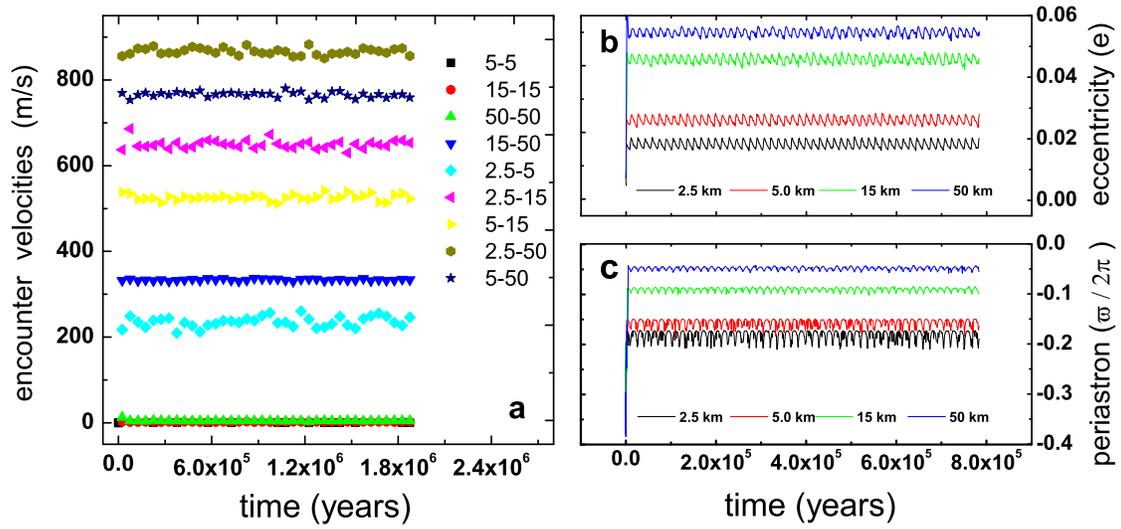}
  \caption{Same with figure 4 but for constant gas case}
\end{figure}

\clearpage
\begin{figure}
\includegraphics[width=\textwidth]{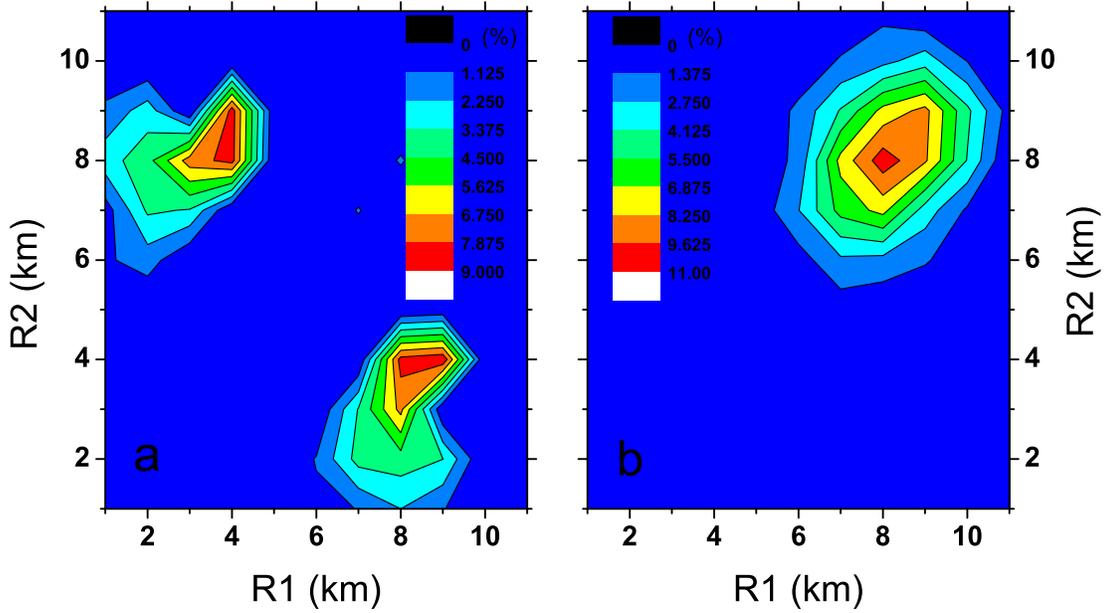}
  \caption{Distributions of impact rates in R1-R2 plane.
  R1, R2 are the radiuses of the two colliding bodies.
  The impact rates are computed as percentages of impacts that occur in areas of a given size in the R1-R2 plane.
  a: the case similar to that in figure 4, in which gas drag and companion's perturbations are included.
  b: a case for compare, in which planetesimal eccentricities and periastrons are
  random, and gas drag and companion's perturbations are not include(otherwise orbital elements will not be random any more)
  For both cases, a Gaussian size distribution, centered on $R_{p}=$8km, is assumed for the size distribution of planetesimals.  }
\end{figure}

\clearpage
\begin{figure}
\includegraphics[width=\textwidth]{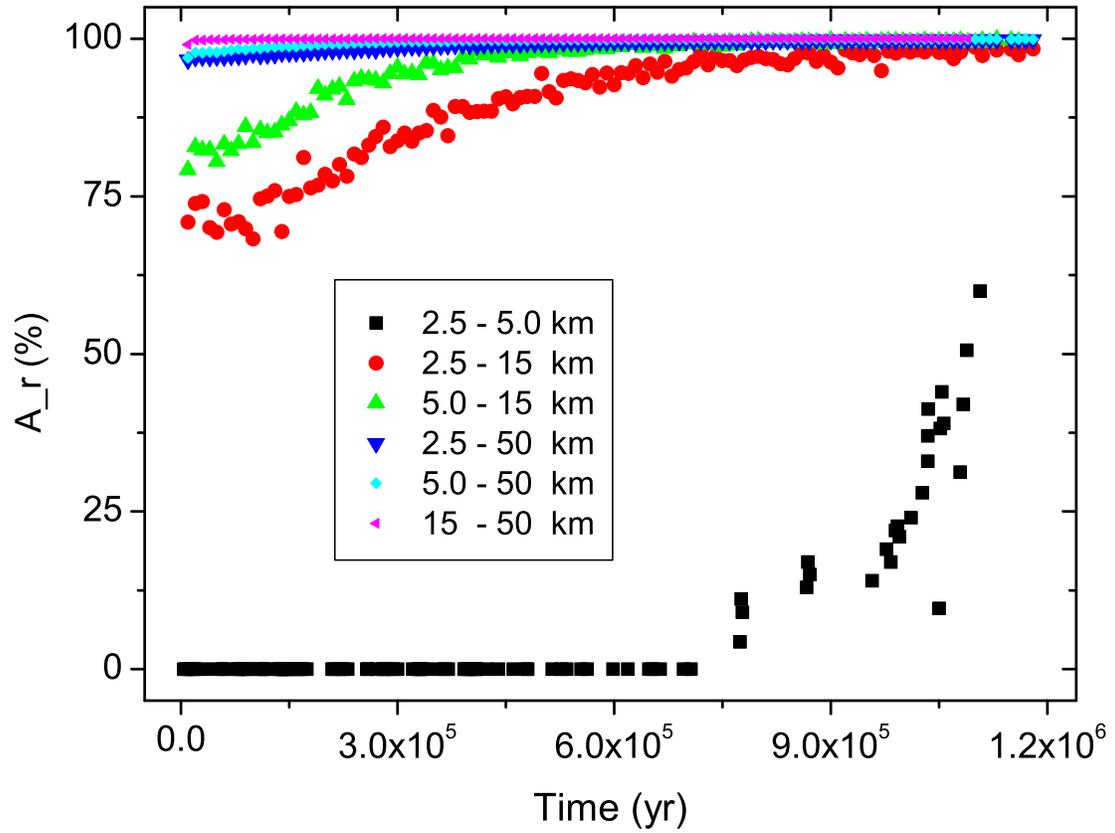}
  \caption{Time-evolutions of net mass accretion ratios for impacts between bodies of different sizes.
  Different types of impacts are plotted in different styles.
  The net mass accretion rates($A_{r}$) are defined and computed following the procedure described in the Appendix}
\end{figure}

\clearpage
\begin{figure}
\includegraphics[width=\textwidth]{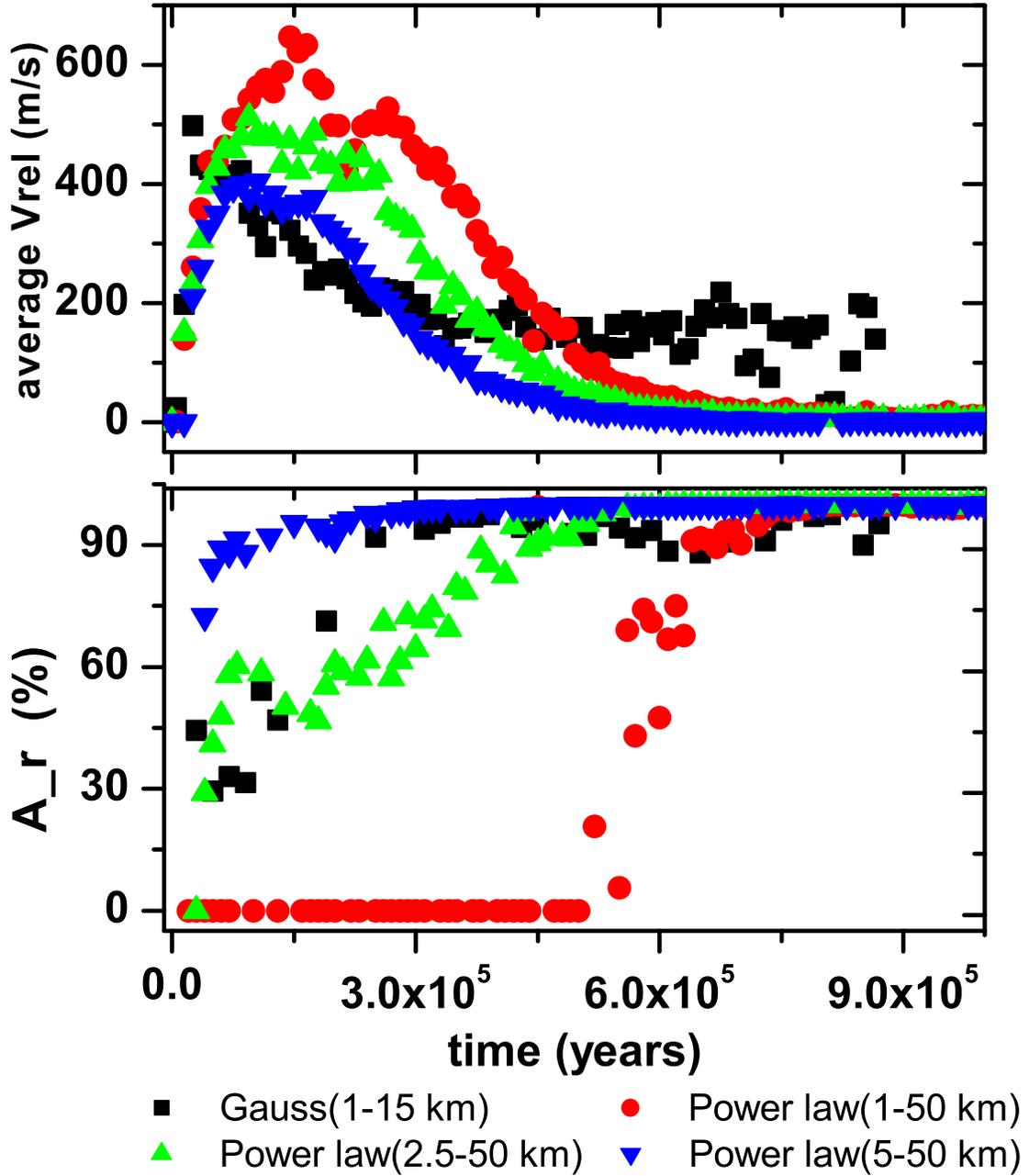}
  \caption{Correlation between net mass accretion ratio and the average relative velocity ($\triangle
  V$), a: time evolution of $\triangle V$, b: time evolution of net mass accretion ratio.
Cases with different size distributions are plotted in different styles.
  The net mass accretion rates($A_{r}$) are defined and computed following the procedure described in the Appendix }
\end{figure}

\clearpage
\begin{figure}
\includegraphics[width=\textwidth]{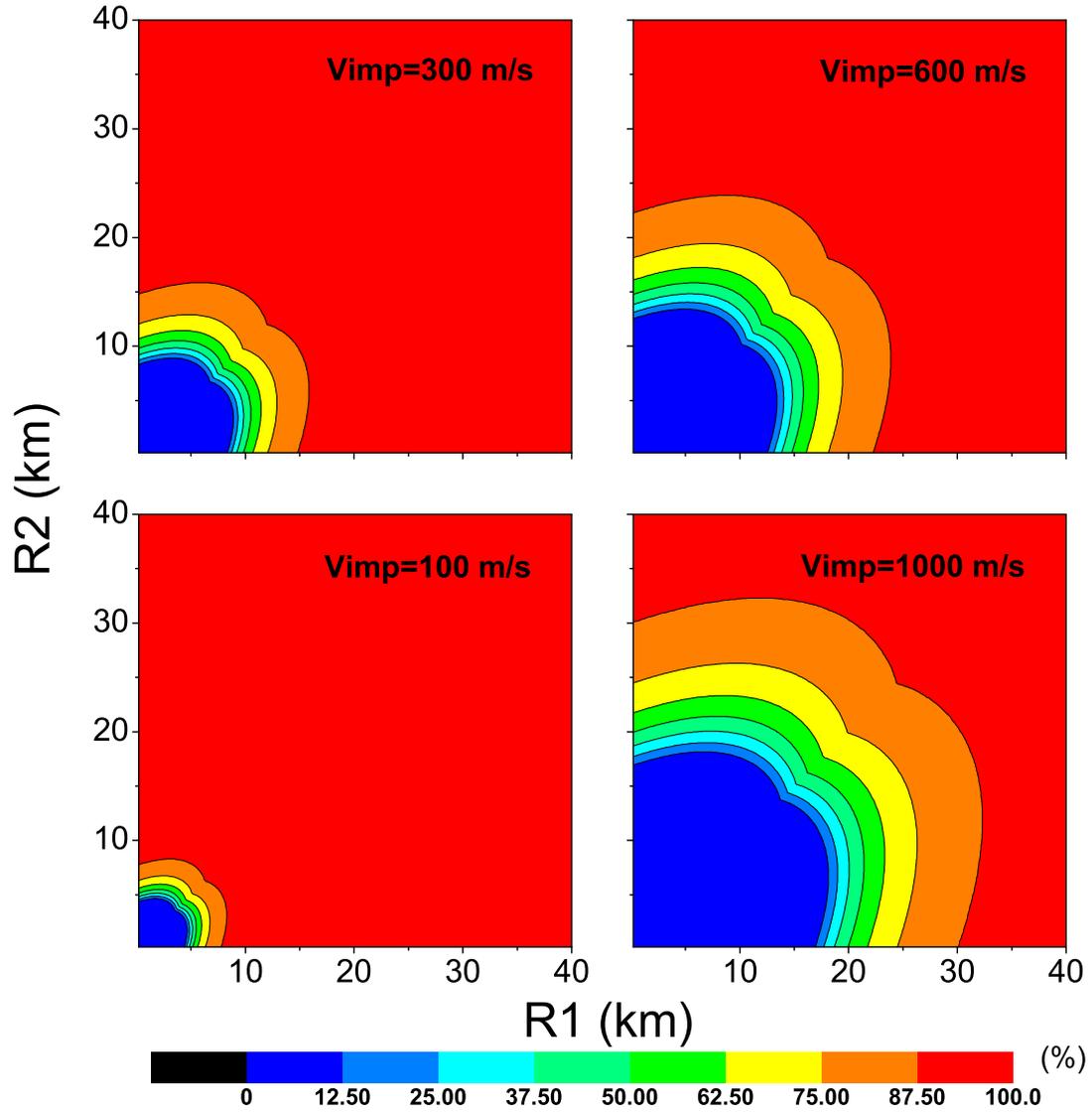}
  \caption{Net mass accretion ratios distributed in R1-R2 plane for 4 typical relative velocities: 100m/s, 300m/s, 600m/s, and 1000m/s.
     R1, R2 are the radiuses of the two colliding bodies. The net mass accretion rates are defined and computed following the procedure described in the Appendix}
\end{figure}

\clearpage



\begin{table}
 \begin{center}
  \caption{Initial Parameters for Runs.\label{tbl-1}}
  \begin{tabular}{@{}llllllllllrrr@{}}
\tableline\tableline
 Paremeters & & & Values & & & Units & & & Descriptions \\
 \tableline
 $a_{B}$ & & & 18.5 & & & $AU$ & & & semi-major axis of companion \\
 $e_{B}$ & & & $0.361\pm 0.023$ & & & - & & & eccentricity of binary\\
 $M_{A}$ & & & 1.6 & & & $M_{\odot}$ & & & mass of primary \\
 $M_{B}$ & & & 0.4 & & & $M_{\odot}$ & & & mass of companion \\
 $R_{p}$ & & & 2.5,5,15,50 & & & $km$ & & & physical radius of planetesimals \\
 e & & & $0\sim0.1$ & & & - & & & initial eccentricities of planetesimals \\
 i & & & $0\sim10^{-3}$ & & & - & & & initial inclinations of planetesimals \\
 $\rho_{p}$ & & & 2.0 & & & $g\cdot cm^{-3}$ & & & planetesimal density \\
 $\rho_{g20}$ & & & $2\times 10^{-9}$ & & & $g\cdot cm^{-3}$ & & & initial gas density at 2AU \\
 $\rho_{g}$ & & & $r^{-2.75}$ & & & - & & & initial gas density radial profile \\
\tableline
\end{tabular}
\tablecomments{$M_{\odot}$ stands for the solar mass}
\end{center}
\end{table}



\end{document}